\documentclass[preprint,showpacs,preprintnumbers,superscriptaddress,floatfix]{revtex4}
\usepackage{graphicx}
\usepackage{dcolumn}
\usepackage{bm}
\usepackage{graphics}

\begin{document}

\title{Monte Carlo results for the hydrogen Hugoniot}

\author{V. Bezkrovniy}
\affiliation{Institut f\"ur Physik, Ernst-Moritz-Arndt-Universit\"at Greifswald, Domstrasse 10a, D-17487, Greifswald, Germany}
\author{V. S. Filinov}
\affiliation{Institut for High Energy Density, Russian Academy of Science, Izhorskaja str. 13/19, 125412 Moscow, Russia}
\author{D. Kremp}
\affiliation{Fachbereich Physik, Universit\"at Rostock, Universit\"atsplatz 3, D-18051 Rostock, Germany}
\author{M. Bonitz}
\affiliation{Christian-Albrechts-Universit\"at zu Kiel, Institut f\"ur Theoretische Physik und Astrophysik, Leibnizstr. 15, 24098 Kiel, Germany}
\author{M. Schlanges}
\affiliation{Institut f\"ur Physik, Ernst-Moritz-Arndt-Universit\"at Greifswald, Domstrasse 10a, D-17487, Greifswald, Germany}
\author{\\ W. D. Kraeft}
\affiliation{Institut f\"ur Physik, Ernst-Moritz-Arndt-Universit\"at Greifswald, Domstrasse 10a, D-17487, Greifswald, Germany}
\author{P. R. Levashov}
\affiliation{Institut for High Energy Density, Russian Academy of Science, Izhorskaja str. 13/19, 125412 Moscow, Russia}
\author{V. E. Fortov}
\affiliation{Institut for High Energy Density, Russian Academy of Science, Izhorskaja str. 13/19, 125412 Moscow, Russia}

\date{\today}

\begin{abstract}
We propose a theoretical Hugoniot obtained by combining results for the equation of state (EOS) from the Direct Path
Integral Monte Carlo technique (DPIMC) and those from Reaction Ensemble Monte Carlo (REMC) simulations.
The main idea of such proposal is based on the fact that DPMIC provides first-principle results  for a wide range
of densities and temperatures including the region of partially ionized plasmas. On the other hand, for lower temperatures
where the formation of molecules becomes dominant, DPIMC simulations become cumbersome and inefficient. For this region
it is possible to use accurate REMC simulations  where bound states (molecules) are treated on the Born-Oppenheimer level using
a binding potential calculated by Kolos and  Wolniewicz.  The remaining interaction is then reduced to the scattering between neutral
particles which is reliably treated classically applying effective potentials. The resulting Hugoniot is located between the experimental
values of Knudson {\textit{et al.}} \cite{1} and Collins {\textit{et al.}} \cite{2}.
\end{abstract}

\maketitle

The H-plasma is a very important and interesting many particle system. Hydrogen is the simplest and at the  same time the most
abundant element in the universe. Due to its high relevance for modern astrophysics, inertial confinement fusion and fundamental
understanding of condensed matter, hydrogen continues to be actively studied both, experimentally \cite{1,2,3,4,5,6} and theoretically \cite{7,8,9,10,11,12,13,14}.
At high temperatures and pressures, the hydrogen behavior is defined by the interaction between free electrons and
protons (plasma state). With decreasing temperature, the contribution of bound states such as atoms and molecules
to the EOS of hydrogen becomes of increasing importance, which at low temperatures completely define the hydrogen properties.
Furthermore, as pointed out in many papers (Norman and Starostin \cite{15}, Ebeling {\textit{et al.}} \cite{16},
Haronska {\textit{et al.}} \cite{17}, Saumon and Chabrier \cite{18}) there are strong theoretical arguments
for a phase transition between two plasma phases. This issue which is of importance, for example,  for models of  Jovian
planets is still actively debated. Among other important questions we mention the high-pressure compressibility, details
of the pressure ionization and dissociation.

For this reason, in the last decades considerable experimental and theoretical investigations were carried out to accurately determine the EOS of hydrogen at high pressures. Experimentally, the EOS for this region can be obtained
using shock-wave techniques. The results of these experiments are usually discussed in form of an Hugoniot
\begin{equation}\label{p1}
E = E_{0}+\frac{1}{2}(p+p_{0})\left(\frac{1}{\rho}-\frac{1}{\rho_{0}}\right),
\end{equation}
where the specific internal energy $E$ at a state with the density $\rho$ and the pressure $p$ is connected to
the initial conditions with the density $\rho_{0}$, the pressure $p_{0}$ and the internal energy $E_{0}$.

\begin{table*}
\caption{Thermodynamic properties of deuterium plasma calculated by DPIMC}
\begin{ruledtabular}
\begin{tabular}{c|cc|cc|cc|c|c}
&\multicolumn{2}{c|}{$r_s=1.7$}& \multicolumn{2}{c|}{$r_s=1.86$}& \multicolumn{2}{c|}{$r_s=2$}  &   &  \\
$T$, K                  & $P$, GPa   & $E$, eV & $P$, GPa    & $E$, eV      & $P$, GPa    & $E$, eV  & $ \rho_H$, g/$cm^3$ & $P_H$, GPa \\ \hline
15625                   & 227.01     & -18.9953 & 101.41     & -9.6854       &                 &               & 0.8539           & 111.32 \\
31250                    & 186.25     & -9.94854 &                &   & 134.30   & -6.0186                     & 0.8370           & 160.53 \\
62500                    &  &         & 314.11     & -1.2281   & 261.05         & -0.1776          & 0.8104     & 306.69 \\
$1.25\cdot 10^5$   &  &         & 7.9579    & 1727.41   & 6.2214        & 0.7395           & 0.7395      &  700.75\\
$2.5\cdot 10^5$     &  &         & 1596.84  & 48.2211   & 1237.67      & 46.8531          & 0.7204     & 1330.47 \\
$5\cdot 10^5$        &  &         & 3261.65  & 112.7294 & 2645.01      & 114.5706        & 0.7082      & 2797.26 \\
$10^6$                 &  &         & 6765.75  & 245.9921 & 5439.83      & 246.4489       & 0.6979      & 5672.16 \\
\end{tabular}
\end{ruledtabular}
\end{table*}

One of the well established experimental techniques for the creation of shock waves uses gas gun devices. With gas gun
experiments, Nellis {\textit{et al.}} \cite{3} reached maximum pressures of 20 GPa and temperatures of 7000 K.
More advanced techniques, the laser-driven experiments used by Collins {\textit{et al.}} \cite{2} and Da Silva
{\textit{et al.}} \cite{4}, allow to reach pressures up to 300 GPa. At such pressures, as expected, hydrogen
transforms from a molecular to a metallic state \cite{5}. The results of laser-driven experiments have shown an
unusual high compression $\rho/\rho_{0}=6$ of deuterium, which deviates significantly from a maximum compression of
$\rho/\rho_{0}= 4.4$  obtained within the SESAME EOS \cite{19}.  However, the experiments of Knudson {\textit{et al.}}  \cite{1}
which used magnetically driven flyer techniques (Z-pinch) do not support such high compressibilities and are close
to those of SESAME \cite{19} and Restricted Path Integral Monte Carlo (RPIMC) \cite{7} results.
The reason for this discrepancy of the two experiments is not yet completely understood and requires more
detailed study \cite{20}, including independent theoretical investigations which is the aim of this paper.
It is also necessary to mention other important experimental techniques such as
the {\it convergent geometry} technique \cite{6}. The experimental point obtained by Belov {\textit{et al.}} \cite{6}
within this technique is located between the results of $\it laser-driven$ and $\it magnetically~ driven ~flyer$ experiments.

An Hugoniot can be also determined theoretically from the equation of state. This enables us to compare
different theoretical approaches and computer simulations with experimental results, which cover
a large region in the phase diagram of hydrogen. They start at temperatures of about $20$~$\textrm  K $ and at a density
of $\rho_{0}=0.171$~${\textrm g/cm^{3}}$, which corresponds to the liquid state, and go up to
temperatures and densities where only free electrons and nuclei exist. To our knowledge, there is no theory or
computer simulation which rigorously and consistently describes the complete region of the EOS achievable by
experiments. For example, the linear mixing model (LM) of Ross \cite{21} rather well predicts the behavior of the laser
driven experiments; however it is a semi--empirical theory which interpolates between molecular and metallic states
of hydrogen.

\begin{figure}[b]
\includegraphics[width=8cm,height=8cm,clip=true]{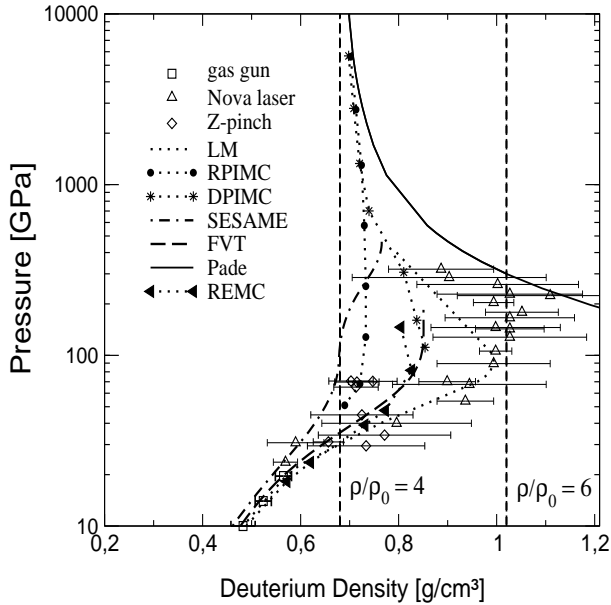}
\caption{Experimental and theoretical results for the deuterium Hugoniot}
\label{fig1}
\end{figure}

Further, the region of completely and partially ionized hydrogen can be described analytically using the methods of
quantum statistics \cite{16,22,23}. In such methods, an EOS is obtained from a fugacity
expansion (ACTEX) \cite{23} and  modified fugacity expansions which are upgraded by means of quantum--field
theoretical methods (leading to dynamical screening, self energy and lowering of the ionization energy \cite{16,22}).
In the latter case it is useful to condense the results in form of Pad\'e approximations \cite{24}, (from Debye to
Gellman--Brueckner). Of course, the EOS following  from these theories cannot reproduce the Hugoniot over the
full range of density and pressure. It gives only the asymptotic behavior at higher temperatures.
The typical behavior of the analytical theory \cite{24} is shown in Fig.~\ref{fig1}. It coincides only
asymptotically with the {\it ab initio} RPIMC calculations and, with decreasing temperature, deviates
considerably from those results. The Hugoniot calculated within the ACTEX theory which is not shown here
exhibits a similar behavior \cite{23}.

The main reason for the failure of the  analytical  theories is obvious. As we mentioned already, for lower temperatures, the neutral particles, i.e., H-atoms and H$_{2}-$molecules, become increasingly important, giving rise to a
strongly coupled dense gas or liquid. Under such conditions it is necessary to invoke the methods of  the theory of liquids.
The simplest theory for this region is the fugacity expansion of the EOS up to the second virial coefficient \cite{25}.
This theory is applicable only for low densities and cannot correctly describe  the molecular dissociation
which is an important process occurring during shock wave experiments. For moderate densities,  the fluid variational theory (FVT),
proposed by Ross {\textit{et al.}} \cite{26}, can be applied.  This theory was further developed by Juranek and Redmer \cite{12}
to many component systems, where molecular dissociation occurs. The effective interaction potentials \cite{26,27} between components
used within FVT achieve good agreement with experimental gas gun data of Nellis {\textit{et al.}} \cite{3}, Fig. \ref{fig1}.

A powerful tool for the investigation of the hydrogen EOS is {\it ab initio} computer simulation. Quantum
molecular dynamics simulations, based on a density functional theory, are usually applied to investigate the
atomic and molecular region \cite{8,13} but have difficulties to describe the partially ionized plasma.
The wave packet molecular dynamics also covers the region of the fully ionized plasma \cite{9} but yields
unexpectedly high compressibilties. In this work we will not discuss these methods in detail and refer to the work cited.

The {\it Path Integral Monte Carlo} method is another first principle method which is well suited for the
investigation of the EOS of hydrogen over a wide parameter range. Except for the problem of the Fermi statistics,
it is an exact solution of the many-body quantum problem for a finite system in thermodynamic equilibrium.
The treatment of the ``sign problem'' makes the main difference between the RPIMC method used by Ceperley and Militzer \cite{7}
and the Direct Path Integral Monte Carlo (DPIMC) method used by Filinov {\textit{et al.}} \cite{10,11} and others. This problem is beyond the present paper, here we
restrict ourselves to the discussion of the DPIMC method.

The idea of DPIMC is the well known: any thermodynamic property of a two-component plasma with $N_e$ electrons
and $N_p$ protons at a temperature $T$ and volume $V$ is defined by the partition function $Z(N_e, N_p, V, T)$:
\begin{equation}
Z(N_e, N_p, V, T)=\frac{1}{N_e!N_p!}\sum_{\sigma}\int_{V}{%
dq\,dr\,\rho(q,r,\sigma;T)},
\end{equation}
where $q$ ($r$) comprises the coordinates of the protons (electrons), $\sigma $ stands for the spin of the electrons, and $\rho $ is the density matrix of the system. Taking into account
the electron spin and the Fermi statistics (antisymmetrization), the density matrix  is expressed by a path
integral \cite{28} where all electrons are represented by fermionic loops with a
number of intermediate coordinates (beads). In our simulations, we used an effective quantum pair potential, which is finite at zero
distance \cite{29}. This potential was obtained by Kelbg as a result of a first--order perturbation theory.
The simulation has been performed at temperatures of $10^4~ {\textrm K}$ and higher in a wide range of particle densities.
Under these conditions the exchange effects for protons are negligible. In the present calculations,
we used an improved treatment of the electron exchange, i.e., we took into account the exchange interaction of electrons
from neighbor Monte Carlo cells, namely first from the nearest neighbor cells ($3^3-1$), then from the next
neighbors ($5^3-1$) and so on. The calculated thermodynamic properties of hydrogen allowed us to compute the shock
Hugoniot of deuterium using  Eq.~(\ref{p2})

\begin{equation}\label{p2}
H = E -  E_{0} - \frac{1}{2}(p+p_{0})(V-V_0) = 0.
\end{equation}

\begin{table}[t]
\caption{Hugoniot data calculated by REMC}
\begin{ruledtabular}
\begin{tabular}{l|l|l|l|l|l|l|l}
$T$, K                   & 2000    &  4000    &   5000   &   8000  & 10000  & 13000  & 15000  \\
$\rho$,  g/$cm^3$      & 0.470   &  0.570   &   0.618  &   0.729 &  0.771  & 0.804   &  0.815  \\
$P$, GPa             & 9.183   &18.690   &  23.96   &  39.35  & 47.823 & 58.71   &  65.43
\end{tabular}
\end{ruledtabular}
\end{table}
Following the work \cite{7} we chose $p_{0}=0$, $\rho_0=0.171$ g/cm$^3$, $E_0=-15.886$ eV per atom and computed the pressure $p_i$ and the energy $E_i$ at a given constant temperature $T$ (from $10^4$ K to $10^6$ K) and three values of the
volume $V_i =1/\rho_i$ corresponding to $r_s=$1.7, 1.86, and 2, where $r_s=\bar r/a_B$, $\bar r=(3/4\pi n_p)^{1/3}$, $n_p$ is the particle density, $a_B$-- Bohr radius. The results of the calculations are presented in Table 1. Substituting the obtained values $p_i$, $E_i$ and $V_i$
into the Hugoniot we determine the volume range $V_1, V_2$ where the function $H(p,V,E)$ changes its sign. The
value of the density at the Hugoniot is calculated by linear interpolation of the function $H$ between $V_1$ and
$V_2$. The values of the pressure and of the total energy are shown in the Table 1 only for those density values
between which the value of the density lies on the Hugoniot at a given temperature. The values
of density and pressure on the Hugoniot are placed in the last two columns of Table 1.
and are polotted together with selected theoretical and experimental data in Fig.~\ref{fig1}.
The lowest temperature included in this figure for the DPIMC is 15625 K.

In order to correctly describe the quantum mechanics of the formation of molecules at temperatures lower than
$10000~{\textrm  K}$, it is necessary to take many beads. In this region, DPIMC calculations become very time consuming
and the convergence is poor. The natural proposal which appears for this region is to use the asymptotic property
of the path integral which, for heavy particles, goes over into the classical partition function.
For such systems, the classical Monte Carlo scheme can be applied. An advanced version of the classical Monte Carlo
scheme is the reaction ensemble Monte Carlo technique (REMC) \cite{30}. This method incorporates the quantum mechanical
description of bound states, while the scattering states are treated classically. As was shown by Bezkrovniy
{\textit{et al.}} \cite{14}, REMC describes the low temperature region very well, and yields good agreement
with the gas gun experiments by Nellis {\textit{et al.}} \cite{3} Fig.~\ref {fig1}.  In these simulations the energy levels for the
molecular partition functions of hydrogen and deuterium are obtained by solving the Schr\"odinger equation with the potential
calculated by Kolos and Wolniewicz \cite{31}.   On the basis of the REMC, results are obtained much easier as compared to those
from molecular  dynamics based on density functional theory; see Bonev {\textit{et al.}} \cite{13} and Fig. 2.
Our REMC data are presented in Table 2.
\begin{figure}
\includegraphics[width=8cm,height=8cm,clip=true]{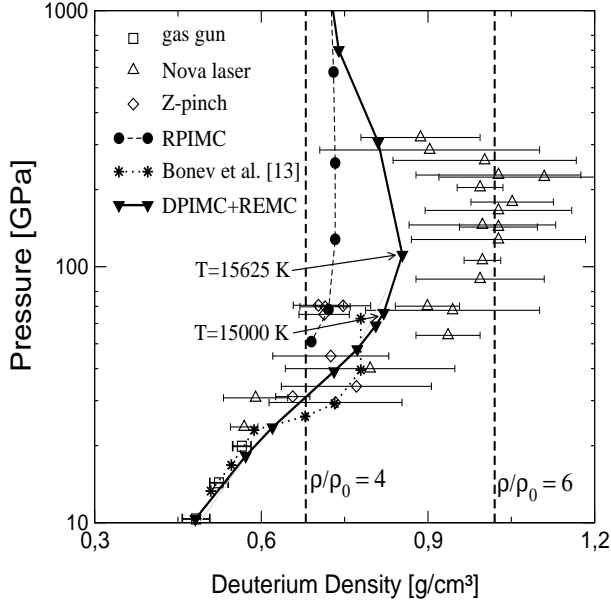}
\caption{Results for the combined Hugoniot}
\label{fig2}
\end{figure}

In order to get a unified picture combining DPIMC and REMC, we use the fact that REMC  turns out to be the
limiting case of DPIMC at low temperatures, where hydrogen consists only of atoms and molecules.
Therefore, it is obvious to use the asymptotic results of both methods to construct an Hugoniot which can  be applied in the entire range of compression. For the construction of the combined Hugoniot we carefully analyzed the region where the Hugoniots produced by the two methods can be connected to each other. As we can see from Fig. \ref{fig1} the Hugoniot calculated within DPIMC ends at the point $15625~{\textrm K}$.
At this temperature, the largest contribution to the EOS are given by molecular states. As natural continuation of the
DPIMC Hugoniot, we take the point of $15000~{\textrm K}$ produced by REMC. We want to stress here that these two methods are completely independent and no interpolation procedure is used. Just two points at $15625~{\textrm K}$ of DPIMC and $15000~{\textrm  K}$
of REMC are connected to each other. The final Hugoniot is  plotted in Fig. \ref{fig2} and shows a maximum compressibility of
approximately 4.75 as compared to the initial deuterium density.

\begin{acknowledgments}
The authors greatfully acknowledge fruitful discussions with R. Redmer, Th. Bornath, and H. Juranek (Rostock).
The work was supported by the Deutsche Forschungsgemeinschaft (SFB 198, BO 1366/2), the grant for talented young
researchers of the Science  support foundation, Rus. Fed. President Grant No. MK-1769.2003.08, the RAS program No. 17
``Parallel calculations and multiprocessor computational systems'', grant of the U.S.
Civilian Research
and Development Foundation for Independent States of the Former Soviet Union (CRDF) No. PZ-013-02
and the Ministry of Education of Russian Federation and by a grant  for CPU time at the NIC  J\"ulich.
\end{acknowledgments}

\end{document}